\documentclass[aps,prl,twocolumn,superscriptaddress,showpacs]{revtex4}
\usepackage{tipa}
\usepackage{pifont}
\usepackage{txfonts}
\usepackage{amssymb}
\usepackage{graphicx}

\begin{document}

\title{Conservation laws of the Haldane-Shastry type spin chains}
\author{Junpeng Cao}
\affiliation{Beijing National Laboratory for Condensed Matter
Physics, Institute of Physics, Chinese Academy of Sciences, Beijing
100190, People's Republic of China}
\author{Peng He}
\affiliation{Beijing National Laboratory for Condensed Matter
Physics, Institute of Physics, Chinese Academy of Sciences, Beijing
100190, People's Republic of China}
\author{Yuzhu Jiang}
\affiliation{Beijing National Laboratory for Condensed Matter
Physics, Institute of Physics, Chinese Academy of Sciences, Beijing
100190, People's Republic of China}
\author{Yupeng Wang$^*$}
\affiliation{Beijing National Laboratory for Condensed Matter
Physics, Institute of Physics, Chinese Academy of Sciences, Beijing
100190, People's Republic of China}

\begin{abstract}
A systematic method to construct the complete set of conserved
quantities of the Haldane-Shastry type spin chains is proposed. The
hidden relationship between the Yang-Baxter relation and the
conservation laws of the long-range interacting integrable models is
exposed explicitly. An integrable anisotropic Haldane-Shastry model
is also constructed.
\end{abstract}

\date{\today}
\maketitle


A typical characteristic of the integrable models is that each of
them possesses a complete set of conservation laws. For an
integrable system with a finite number of degrees of freedom, the
number of linearly independent conserved quantities is exactly the
same as that of the degrees of freedom, while for a continuous
integrable model, there is an infinite number of conserved
quantities. It is well known that the conservation laws of the
integrable models with short-range interactions are tightly related
to the Yang-Baxter relation. The conserved quantities can be
obtained from the derivatives of the transfer matrix of the system
\cite{tak,bax}. On the other hand, there is another class of
integrable models with $r^{-2}$ type interaction potentials which
are called as the Calogero-Sutherland (CS) model \cite{cs} in the
continuous case and the Haldane-Shastry (HS) model \cite{hs-1,hs-2}
in the lattice case. These models belong to the long-range
interacting ones and have many applications in the fields of
two-dimensional fractional quantum Hall effect and fractional
statistics \cite{st}. Because of the importance, these models have
been studied extensively
\cite{hs2,h21,h12,h2,s2,hs3,hs4,hs5,hs6,hs7,hs8}. Several people
have tried to construct the conservation laws of these models. For
example, by using the Dunkl operators \cite{dun}, Polychronakos
\cite{pol1} constructed the invariants of motion of the CS
continuous model. Nevertheless, the demonstration of the
integrability of the HS Lattice model is still a challenge problem.
Borrowed Polychronakos's idea, Fowler and Minahan \cite{fow}
proposed a set of conserved quantities for the HS model. However,
there are still some debates \cite{cyb,pol2} on this construction
because the HS Hamiltonian appears in the third level of invariants,
and should act on some magnon states to erase the unwanted terms.

The main difficulty of the demonstration of the complete
integrability of HS type models is lack of a systematic method to
construct the complete set of conservation laws. Recalling the case
of nearest neighbor interacting integrable models,  an interesting
question may arise: Is there any intrinsic relationship between the
HS model and the Yang-Baxter relation \cite{cyb}? In this Letter, we
expose this intrinsic relationship exactly and develop a systematic
method to construct the complete set of conservation laws of HS
model. This provides us not only a deep understanding to this
important model but also a general method to construct other
integrable models with long-range couplings.

Our starting point is the following Lax operator
\begin{eqnarray}
L_{0j}(u)=1+\frac{\eta}u{\bf \sigma}_0\cdot{\bf S}_j, \label{hl-1}
\end{eqnarray}
where $u$ is the spectral parameter; $\eta$ is the crossing
parameter; ${\bf \sigma}_0$ is the auxiliary Pauli matrix  and ${\bf
S}_j$ is the spin-$1/2$ operator on site $j$. It is well known that
the integrability of the Heisenberg spin chain model is related to
the monodromy matrix $T_0(u) = L_{01}(u)\cdots L_{0N}(u)$ which
satisfies the Yang-Baxter relation
\begin{equation}
L_{12}(u_1-u_2)T_1(u_1)T_2(u_2)=T_2(u_2)T_1(u_1)L_{12}(u_1-u_2).
\label{tybr}
\end{equation}
Define the transfer matrix $t(u)=tr_0 T_0(u)$. From Eq.(\ref{tybr}),
one can prove that the transfer matrices with different spectral
parameters are mutually commutative, i.e., $[t(u), t(v)]=0$.
Therefore, $t(u)$ serves as the generating function of the conserved
quantities of the corresponding system. The first order derivative
of logarithm of the transfer matrix gives the Hamiltonian of the
Heisenberg spin chain
\begin{eqnarray}
H_{H}=\frac{1}{2\eta} \left.\frac{\partial }{\partial u}\ln t(u)
\right|_{u=\frac \eta 2}+(\frac 1 {\eta^2}- \frac 14)N=\sum_{j=1}^N
{\bf S}_j \cdot {\bf S}_{j+1}.
\end{eqnarray}

In fact, the inhomogeneous Lax operator $L_{0j}(u-\delta_j)$ with a
site-dependent shift $\delta_j$ to the spectral parameter $u$ and
the inhomogeneous transfer matrix $\tilde T_{0}(u)\equiv
L_{01}(u-\delta_1)\cdots L_{0N}(u-\delta_N)$ also satisfy the
Yang-Baxter relation Eq.(\ref{tybr}). Consider the classical
expansion of the Lax operator $L_{0j}(u-\delta_j)=1+\eta {\cal
L}_{0j}(u-\delta_j)$ and define the classical monodromy matrix
${\cal T}_0(u) = \sum_{j=1}^N {\cal L}_{0j}(u-\delta_j)$. We find
that they satisfy the following classical Yang-Baxter relation
\cite{skly}
\begin{equation}
[{\cal T}_{1}(u_1), {\cal T}_{2}(u_2)]= [{\cal T}_{1}(u_1)+{\cal
T}_{2}(u_2), {\cal L}_{1 2}(u_1-u_2)]. \label{cybr}
\end{equation}
The Eq.(\ref{cybr}) ensures that the functional $\tau(u)\equiv tr_0
{{\cal T}_0}^2(u)/4$ with different spectral parameters are mutually
commutative, i.e., $[{\tau}(u), {\tau}(v)]=0$. Therefore,
${\tau}(u)$ can be treated as the generating functional of a series
of conserved quantities which can be written out explicitly as
\begin{equation}
{\tau}(u)=\sum_{j=1}^{N}\frac{3}{8(u-\delta_{j})^2} +
\sum_{j=1}^{N}\frac{h_j}{u-\delta_j},
\end{equation}
where
\begin{eqnarray}
&&h_j=\sum_{l=1, \neq j}^{N} \frac{1}{\delta_{j}-\delta_{l}} {\bf
S}_j \cdot {\bf S}_{l}.\label{rgau}
\end{eqnarray}
The operators $h_j$ are nothing but the Gaudin operators \cite{gau}
associated with the Heisenberg spin chain. It is easy to prove that
the Gaudin operators commute with each other, i.e., $[h_j, h_k]=0$.
This allows us to construct the  mutually commutative operators
$I_n=\sum_{j=1}^N {h_j}^n$ for arbitrary $n$. If we choose one of
$I_n$ as the Hamiltonian, $\{I_n\}$ form a set of conserved
quantities and the Hamiltonian describes an integrable system. For a
translational invariant lattice $\delta_j= j$ and $N\to\infty$, we
find
\begin{eqnarray}
H_{ISE}&=&\lim_{N\to\infty}\left[-\sum_{j=1}^Nh_j^2+\frac{\pi^2}{16}N\right]\nonumber\\
&=&{\sum_{l<j}}^\prime \frac{1}{(j-l)^2}{\bf S}_j \cdot {\bf S}_l.
\label{yy1}
\end{eqnarray}
$H_{ISE}$ is just the Hamiltonian of inverse square exchanging (ISE)
model. The prime in the summation means that $j$ and $l$ in the
summation take values from negative to positive infinity. Obviously,
operators $h_j=\sum_{l\neq j}^\prime(j-l)^{-1}{\bf S}_j \cdot {\bf
S}_{l}$ and their arbitrary combinations commute with the
Hamiltonian (\ref {yy1}) and form a set of the conserved quantities.
In addition, the model (\ref {yy1}) has another set of conserved
quantities $h_j'=\sum_{l\neq j}^\prime{\bf S}_j \cdot {\bf S}_{l}$
by simply checking $[H_S, h_j']=0$. We note that (\ref{yy1}) with
arbitrary $\delta_j$ gives a disordered integrable system. Now it is
clear that there is an intrinsic relationship between the
short-range coupling Heisenberg model and the long-range coupling
ISE model, i.e., they share the common Yang-Baxter equation. This
provides us a powerful method to construct new integrable models
with long-range interactions from the known solutions of the
Yang-Baxter equation or to obtain the conservation laws of the
predicted integrable models with $r^{-2}$ type potentials. For
example, from the Lax operator of the anisotropic XXZ Heisenberg
spin chain \cite{tak} we have the following mutually commutative
Gaudin operators
\begin{eqnarray}
h_j=\sum_{l=1,\neq j}^N\left[
\frac{S_j^xS_l^x+S_j^yS_l^y}{\sin(\delta_j-\delta_l)}+\cot(\delta_j-\delta_l)
S_j^zS_l^z\right]. \label{xxzhj-1}
\end{eqnarray}
For the equally spaced $\delta_j=\pi j/N$,  we obtain
\begin{eqnarray}
H_{AHS}&=&-\sum_{j=1}^N h_j^2-\frac{1}{4} \left(\sum_{j=1}^N
S_j^z\right)^2 +\frac{1}{16}N(N^2-N+1)\nonumber\\
&=&\sum_{l < j}^N
\frac{\cos\frac{\pi}N(j-l)(S_j^xS_l^x+S_j^yS_l^y)+S_j^zS_l^z}{
\sin^2\frac{\pi} N(j-l)}. \label{xxzh-2}
\end{eqnarray}
Notice that in the anisotropic XXZ spin chain, the total spin is no
longer a good quantum number but $\sum_{l=1}^N S_l^z$ is indeed a
conserved quantity which commutes with the $h_j$ in
Eq.(\ref{xxzhj-1}) and the $H_{AHS}$ in Eq.(\ref{xxzh-2}).
Therefore, $H_{AHS}$ can be treated as the Hamiltonian of an
anisotropic HS model. In fact, the ISE Hamiltonian (\ref{yy1}) is
the limiting case of the anisotropic HS model (\ref{xxzh-2}), i.e.,
$H_{ISE}=\lim_{N\to\infty}{\pi^2}/{N^2}H_{AHS}$. By putting
$N\to\infty $ and $\delta_m=i m$, where $i$ is the imaginary unit,
we readily obtain the hyperbolic version of this integrable
Hamiltonian
\begin{eqnarray}
H_{HAHS}={\sum_{l<j}}^\prime
\frac{\cosh(j-l)(S_j^xS_l^x+S_j^yS_l^y)+S_j^zS_l^z}{ \sinh^2(j-l)}.
\end{eqnarray}

Motivated by these findings, we introduce the following local
operators
\begin{eqnarray}
h_j=\sum_{l=1,\neq j}^N f(\delta_j-\delta_l){\bf S}_j \cdot {\bf
S}_{l}, \label{xxzhj}
\end{eqnarray}
and look for the solutions of $[h_j, h_k]=0$, where
$f(\delta_j-\delta_l)$ is a  function  to be determined.  For
simplicity, we denote $f(\delta_j-\delta_l)\equiv f_{jl}$. With the
relation $[{\bf S}_j \cdot {\bf S}_{l}, {\bf S}_j \cdot {\bf
S}_{k}]=i{\bf S}_j \cdot({\bf S}_{l}\times{\bf S}_{k})$ for $l\neq
j\neq k$, we obtain
\begin{eqnarray}
[h_j, h_k]=i\sum_{l=1,\neq j,k}^N
(f_{jl}f_{kj}-f_{jl}f_{kl}+f_{jk}f_{kl}){\bf S}_j \cdot({\bf
S}_{l}\times{\bf S}_{k}).
 \label{ip3}
\end{eqnarray}
Therefore, the constraint for $[ h_j, h_k]=0$ is the solution of
$f_{jl}f_{kj}-f_{jl}f_{kl}+f_{jk}f_{kl}=0$. After some simple
algebra, we find three sets of solutions:

(i) The first solution is $f(x)=x^{-1}$. This solution just gives
Eq.(\ref{rgau}) and thus the ISE Hamiltonian (\ref{yy1}).

(ii) The second solution is $f(x)=\cot(x) \pm i$. The operators
$h_j$ and $h_j^\dagger$ take the forms
\begin{eqnarray}
h_j=\sum_{l=1,\neq j}^N \left[ \cot(\delta_j-\delta_l)+ i\right]
{\bf
S}_j \cdot {\bf S}_l, \nonumber\\
h_j^\dagger=\sum_{l=1,\neq j}^N \left[ \cot(\delta_j-\delta_l)-
i\right] {\bf S}_j \cdot {\bf S}_l. \label{op}
\end{eqnarray}
Both $\{h_j\}$ and $\{h_j^\dagger\}$ are not hermitian for real
$\delta_j$ but each of them form a set of mutually commutative
operators, though $h_j$ and $h_k^\dagger$ do not commute with each
other, $[h_j, h_k^{\dag}]=2\sum_{l=1,\neq j,k}^N \left[
\cot(\delta_j-\delta_k)+i\right] {\bf S}_j \cdot ({\bf S}_l \times
{\bf S}_k)$. A key problem is to construct a hermitian Hamiltonian
from those non-hermitian operators. Fortunately, we find that
$I_2\equiv \sum_{j=1}^N h_j^2=\sum_{j=1}^N{h_j^\dagger}^2$ is a
hermitian operator. For $\delta_j=\pi j/N$, the HS
Hamiltonian\cite{hs-1,hs-2} can be derived as
\begin{eqnarray}
H_{HS}&=&-\sum_{j=1}^N h_j^2 +\frac{i}{2}(N-4)\sum_{j=1}^N h_j
+\frac{1}{16}N(3N^2-6N-5)\nonumber\\
&=&\sum_{j<l}^N \frac{1}{\sin^2\frac{\pi }N(j-l)}{\bf S}_j \cdot
{\bf S}_l.\label{ihs}
\end{eqnarray}
An obvious fact is that $[H_{HS}, h_j]=[H_{HS}, h_j^\dagger]=0$. We
readily have two independent sets of conserved hermitian quantities
for $H_{HS}$:
\begin{eqnarray}
&&{I_j}^+=\frac{1}{2}(h_j+h_j^\dag)=\sum_{l=1,\neq
j}^N \cot\frac{\pi}N(j-l){\bf S}_j \cdot {\bf S}_l, \nonumber \\
&&{I_j}^-=\frac{1}{2i}(h_j-h_j^\dag)=\sum_{l=1,\neq j}^N {\bf S}_j
\cdot {\bf S}_l. \label{opop}
\end{eqnarray}
For the spin half system, we know that the number of  degrees of
freedom of each site is two. The two linearly independent conserved
quantities $I_j^\pm$ clearly show that the HS Hamiltonian is
completely integrable.

Actually, we can also define the classical operators ${\cal
L}_{0j}(u-\delta_j)=[\cot(u-\delta_j)+i]{\bf S}_0 \cdot {\bf S}_j$
and ${\cal T}_0(u)=\sum_{j=1}^N{\cal L}_{0j}(u-\delta_j)$ for this
solution. They satisfy the following deformed classical Yang-Baxter
relation
\begin{eqnarray}
[{\cal T}_1(u), {\cal T}_2(v)]=[{\cal L}_{21}(v-u), {\cal
T}_1(u)]+[{\cal T}_2(v), {\cal L}_{12}(u-v)]. \label{ybr2}
\end{eqnarray}
Define a functional $\tau (u)= tr_0 {\cal T}_0^2(u)$, which can be
written out explicitly as
\begin{eqnarray}
&& {\tau}(u)=\sum_{j=1}^N
\left[\frac{3\cos2(u-\delta_j)}{8\sin^2(u-\delta_j)} + \frac{3}{4}
i\cot(u-\delta_j)\right]\nonumber \\ && \qquad\quad + \sum_{j=1}^N
\left[\cot(u-\delta_j)+i\right] h_j. \label{tau-new}
\end{eqnarray}
From the deformed Yang-Baxter relation (\ref{ybr2}), we can prove
that the functional $\tau(u)$ with different spectral parameters are
mutually commutative and thus can be treated as the generation
functional of a integrable system. Obviously, for $\delta_j=j\pi/
N$, $[H_{HS}, \tau(u)]=0$, implying that the conserved quantities of
the HS model can also be generated by $\tau(u)$.

(iii) The third solution is $f(x)=\coth(x)\pm 1$ and $
h_j^\pm=\sum_{l=1,\neq j}^N \left[ \coth(\delta_j-\delta_l)\pm
1\right] {\bf S}_j \cdot {\bf S}_l$. With the same procedure, we
find that the Inozemtsev Hamiltonian \cite{hs2}
\begin{eqnarray}
H_{I}={\sum_{l<j}}^\prime \frac{1}{ \sinh^2(j-l)}{\bf S}_j \cdot
{\bf S}_l \label{tri2-h}
\end{eqnarray}
can be easily derived from $\sum_j^\prime {h_j^+}^2$ or
$\sum_j^\prime {h_j^-}^2$ by taking $\delta_j =j$ and $N\to\infty$.
The operators $h_j^\pm=\sum_{l\neq j}^\prime \left[ \coth(j-l)\pm
1\right] {\bf S}_j \cdot {\bf S}_l$ span the complete space of the
conserved hermitian quantities and the Inozemtsev model (\ref
{tri2-h}) is also completely integrable.

Similarly, we can construct the conservation laws of the
$SU(M)$-invariant HS and Inozemtsev models. In these cases, the
generating operators $h_j$ take the form:
\begin{eqnarray}
h_j=\sum_{l=1,\neq j}^Nf(\delta_j-\delta_l)(P_{jl}-c),
\end{eqnarray}
where $P_{ij}$ is the $SU(M)$ spin permutation operator and $c$ is a
constant. One can easily check that
\begin{eqnarray}
[h_j, h_k]=\sum_{l=1,\neq j,k}^N
(f_{jl}f_{kj}-f_{jl}f_{kl}+f_{jk}f_{kl})P_{jl}(P_{kj}-P_{kl}).
\end{eqnarray}
The constraint $[h_j, h_k]=0$ gives the same solutions of $f(x)$ as
those in the (i) - (iii). The $SU(M)$ HS Hamiltonian can be
constructed from the linear combination of $\sum_{j=1}^N{I_j^+}^2$
and $\sum_{j=1}^N I_j^+$, whose explicit form reads
\begin{eqnarray}
H_{SU(M)}=\sum_{l<j}^N\frac{P_{jl}}{\sin^2\frac\pi N(j-l)}.
\end{eqnarray}

In conclusion, we establish the intrinsic relationship between the
inverse square potential spin chain models and the Yang-Baxter
relation. This provides us a powerful method to construct new
integrable models with long-range couplings from the solutions of
the Yang-Baxter relation obtained from the systems with short-range
interactions. As an example, an integrable anisotropic HS spin chain
model is derived. The complete sets of conservation laws of the ISE,
HS and Inozemtsev models are constructed in quite simple forms.

This work was supported by the National Natural Science Foundation
of China, the Knowledge Innovation Project of Chinese Academy of
Sciences, and the National Program for Basic Research of MOST.

* Email: yupeng@aphy.iphy.ac.cn

\end{document}